\documentclass[11pt]{article}

\usepackage[margin=1in]{geometry}
\usepackage{setspace}
\usepackage[T1]{fontenc}
\usepackage[utf8]{inputenc}
\usepackage{lmodern}
\usepackage{microtype}

\usepackage{amsmath,amssymb,amsfonts}
\usepackage{mathtools}
\usepackage{bm}
\usepackage{braket}
\usepackage{newtxtext}
\usepackage[varg]{newtxmath}

\usepackage{graphicx}
\usepackage{caption}
\usepackage{subcaption}
\usepackage{booktabs}
\usepackage{array}

\usepackage{csquotes}
\usepackage[
  backend=biber,
  style=phys,
  articletitle=true,
  biblabel=brackets,
  chaptertitle=false,
  pageranges=false,
  doi=false,
  url=false,
  eprint=true,
  isbn=false,
  sorting=none,
  maxbibnames=99
]{biblatex}
\DeclareFieldFormat{titlecase}{#1}
\AtBeginBibliography{\urlstyle{same}}

\usepackage{xcolor}
\usepackage[colorlinks=true,linkcolor=blue,citecolor=blue,urlcolor=blue]{hyperref}

\usepackage{bbold}

\title{\bfseries
Heralded ultrafast generation of macroscopic quantum states in matter with bright squeezed vacuum light}
\author{
Shohei Imai\\
\small Department of Physics, University of Tokyo, Hongo, Tokyo 113-0033, Japan\\
\small \texttt{shohei.imai@phys.s.u-tokyo.ac.jp}\\
\small ORCID: \href{https://orcid.org/0000-0003-3131-2405}{0000-0003-3131-2405}
}
\date{}

\begin{document}
\maketitle

\begin{abstract}
We show that bright squeezed vacuum light, combined with a single-shot quadrature measurement of the post-interaction light, enables the ultrafast generation of macroscopic quantum states in matter.
Although in the weak-coupling regime multiphoton quantum light leaves the unconditional matter state as a classical mixture due to light--matter entanglement, quadrature-based heralding prepares the matter in a Gaussian-weighted quantum superposition of laser-driven matter states.
For an ensemble of resonantly electric-dipole-coupled two-level systems, this heralding dynamics acts as a Gaussian filter with respect to the electric polarization, with brighter squeezed-vacuum light accelerating the preparation of the zero-eigenvalue Dicke state.
Counter-rotating terms further drive a stroboscopic transition from this Dicke state to a cat-like state.
Our results open a route to ultrafast engineering of macroscopic quantum matter with strong-field quantum light.
\end{abstract}

\section{Introduction} \label{sec:introduction}
Macroscopic quantumness lies at the heart of fundamental physics, serving as a bridge between the classical (macroscopic) and quantum (microscopic) worlds and opening new frontiers for future research~\cite{Schrodinger1935, Leggett1980, Frowis2018, Arndt2014}.
Moreover, its ability to surpass classical limits facilitates practical applications in quantum computing~\cite{Shor1994, Grover1996} and quantum metrology~\cite{Helstrom1967, Caves1981, Braunstein1994, Giovannetti2004a}.
However, such macroscopic nonclassicality is intrinsically fragile against interactions with the environment~\cite{Joos1985, Zurek2003}.
This motivates an approach in which macroscopic quantum states are generated on timescales shorter than those of relaxation and dephasing.

Recent advances in strong-field quantum light make this approach increasingly feasible.
Ultrafast quantum optics is developing rapidly~\cite{Stammer2025a, Ciappina2025, Stammer2023, Helversen2023, Lewenstein2024, Cruz-Rodriguez2024a}, driven by the realization of Schr\"{o}dinger cat or kitten states of light through the combination of high-harmonic generation and quantum measurement~\cite{Lewenstein2021, Rivera-Dean2022, Stammer2022a, Stammer2022, Anonymous2025a}, and of bright squeezed vacuum (BSV) light containing as many as $10^{13}$ photons per mode~\cite{Sh.Iskhakov2012, Chekhova2015a, Sharapova2020}.
Correspondingly, studies exploring the interaction between matter and strong-field quantum light have become increasingly active, including the enhancement of nonlinear effects arising from the super-Poissonian photon statistics of BSV~\cite{Spasibko2017a}, high-harmonic generation driven by BSV~\cite{Gorlach2022b, Rasputnyi2024, Wang2024h, Tzur2024, Lemieux2025, Rivera-Dean2025, Gothelf2025}, BSV-driven electron dynamics~\cite{Tzur2022a, EvenTzur2024} and photoemission~\cite{Heimerl2024a, Heimerl2025}, as well as second-harmonic generation~\cite{Lamprou2025} and electron dynamics~\cite{Imai2025b, Imai2026} driven by cat-state light.

However, quantum light gives rise to strong light--matter entanglement.
Consequently, discarding the information carried by the light after the interaction reduces the matter subsystem to a classically mixed state.
To harness the latent quantum character of the matter system, one therefore needs heralding that exploits the information encoded in its entanglement with light.
Indeed, Ref.~\cite{Imai2026} showed that, by performing single-shot projective measurements of photon-number parity or quadrature on the post-interaction light, one can induce a Rabi-oscillation cat state driven by a macroscopic Schr\"{o}dinger cat state of light.

In this paper, we show that single-shot quadrature measurements of the post-interaction BSV light herald the ultrafast preparation of macroscopic quantum states of matter that are inaccessible in the corresponding unconditional dynamics.
We develop a general theory for BSV-driven dynamics by combining the external-field approximation (XFA), which is valid for multiphoton light as well as the realistic weak electron--photon coupling~\cite{Imai2025b, Imai2026}, with the Janszky representation of the squeezed-vacuum (SV) state~\cite{Janszky1990} (Sec.~\ref{sec:general_theory}).
Based on this effective theory, we derive that the quadrature-heralded state of matter is a Gaussian-weighted quantum superposition of matter states driven by laser light [Eq.~\eqref{eq:q-heralded_time_evolution}].

Applying this XFA theory to the Tavis--Cummings (TC) model, the most basic electric-dipole-coupled system~\cite{Tavis1968}, we find that quadrature-heralded matter dynamics is described as an ultrafast Gaussian filter [Eq.~\eqref{eq:time_evolution_XFA}] with respect to the collective electric polarization (or polarization current).
The resulting state is a Dicke state with zero eigenvalue of that macroscopic observable.
Furthermore, when finite measurement resolution is taken into account, the success-probability-weighted quantum Fisher information (QFI) scales as $N^{3/2}$ with the particle number $N$, implying metrological performance beyond the standard quantum limit (SQL) (Sec.~\ref{sec:tavis-cummings:resolution}).
We also analyze the Dicke model~\cite{Rabi1937, Dicke1954} beyond the rotating-wave approximation (RWA) and show that the counter-rotating terms drive a stroboscopic transition from the Dicke state to a $z$-spin-polarized cat-like state (Sec.~\ref{sec:dicke}).
Overall, this work offers a matter-response perspective on a fundamental question in strong-field quantum optics: what is genuinely quantum in the phenomena induced by BSV light?

Unlike previous schemes for quantum state generation using SV light or optical measurements, our framework combines a macroscopic matter system with ultrafast generation enabled by strong fields, a basic electric-dipole coupling, and a single-shot optical readout.
Previous studies using SV light have reported spin squeezing in small systems~\cite{Genes2003}, conditional state generation by continuous quantum nondemolition (QND) optical measurement~\cite{Nielsen2009}, squeezing transfer by complete light absorption~\cite{Kuzmich1997, Hald1999}, long-time steady states under incoherent noise (a broadband bath)~\cite{Agarwal1990}, and quantum-filtering formulations~\cite{Gough2026}.
Other than SV light, proposed schemes have also considered QND-based protocols~\cite{Lemr2009, Stockton2004, Genes2006, Zhang2019h, Massar2003, McConnell2013, McConnell2015, Chen2015b, Vanderbruggen2011, Vanderbruggen2014, Toth2010, Behbood2014, Bouchoule2002}, light--matter disentanglement with quantum erasure~\cite{Leroux2012}, and spin-$1$ ensembles~\cite{Masson2019}.

The remainder of this paper is organized as follows.
Section~\ref{sec:preliminaries} reviews the preliminaries, including the formulation of the XFA for general quantum light (Sec.~\ref{sec:preliminaries:XFA}) and the Janszky representation, which expresses the SV state as a compact superposition of coherent states (Sec.~\ref{sec:preliminaries:Janszky}).
Section~\ref{sec:general_theory} formulates the XFA theory for BSV-driven dynamics and derives general expressions for unconditional states (Sec.~\ref{sec:general_theory:unconditional}) and heralded states (Sec.~\ref{sec:general_theory:heralded}).
Section~\ref{sec:tavis-cummings} applies this theory to the TC model and analyzes optically heralded ultrafast matter dynamics.
It also discusses the measurement resolution and success probability (Sec.~\ref{sec:tavis-cummings:resolution}) and assesses the validity of the XFA by comparison with exact numerical calculations (Sec.~\ref{sec:tavis-cummings:full}).
Section~\ref{sec:dicke} extends the analysis to the Dicke model and examines the effects of the counter-rotating terms.
Section~\ref{sec:conclusion} presents the conclusions.

\section{Preliminaries} \label{sec:preliminaries}

\subsection{External-field approximation (XFA)} \label{sec:preliminaries:XFA}
This section presents a general formulation of the XFA, which is valid when the electron--photon coupling is weak.
Focusing on ultrafast dynamics induced by strong-field quantum light, we analyze the pure-state time evolution of the coupled light--matter system while neglecting relaxation and dephasing.
The idea is to combine the standard external-field treatment of strong-field laser driving with the superposition principle of quantum mechanics.
In the standard treatment, the optical electric field is prescribed as a classical external field, and only the matter dynamics is solved.
This neglects matter-to-light backaction: the coherent state representing the laser is assumed to evolve freely.
Once the quantum optical state is represented as a superposition of coherent states, the linearity of the Schr\"{o}dinger equation allows this external-field treatment to be applied to each coherent-state branch, thereby yielding the XFA effective theory for quantum-light driving.
A detailed derivation using a path-integral formulation and Born-series expansion is given in Ref.~\cite{Imai2026}.

We first represent the initial state of the quantum light as a superposition of coherent states:
\begin{equation}
|\psi(0)\rangle_{\mathrm{p}} = \int \mathrm{d}^2 \alpha\, \chi(\alpha) |\alpha\rangle. \label{eq:def:initial_photon_state}
\end{equation}
Here, $|\alpha\rangle$ ($\alpha \in \mathbb{C}$) denotes a coherent state satisfying $\hat{a}|\alpha\rangle = \alpha|\alpha\rangle$, where $\hat{a}$ is the bosonic annihilation operator.
This property makes coherent states classical states describing laser light.
Because the coherent-state basis is overcomplete, $\chi(\alpha)$ is not unique; the role of this nonuniqueness in the XFA is clarified below in the photonic thermodynamic limit.

Using this representation, the total-system wave function at time $t$ can be approximated as
\begin{equation}
|\Psi_{\mathrm{XFA}}(t)\rangle_{\mathrm{mp}} = \frac{1}{\sqrt{\mathcal{N}}} \int \mathrm{d}^2 \alpha\, \chi(\alpha)|\psi_{F(\alpha)}(t)\rangle_{\mathrm{m}} |\alpha(t) \rangle_{\mathrm{p}}, \label{eq:def:XFA_total-system_wave_function}
\end{equation}
where $\mathcal{N}$ is a normalization constant and $F(\alpha)$ is the field strength; if the electron--photon coupling constant is $g$, then $F(\alpha) = g \alpha$.
The quantity $g$ depends on system-specific details such as the optical mode function and the electric-dipole moment.
The state $|\psi_{F}(t)\rangle_{\mathrm{m}}$ denotes the matter state driven by the classical, laser field $F(\alpha)$, whose time evolution is determined by the following Schr\"{o}dinger equation:
\begin{equation}
\mathrm{i} \partial_t |\psi_{F}(t)\rangle_{\mathrm{m}} = \hat{H}_{\mathrm{m}}[F(\alpha(t))] |\psi_F(t)\rangle_{\mathrm{m}}. \label{eq:def:matter_Schroedinger_eq}
\end{equation}
Here, $\hat{H}_{\mathrm{m}}[F(\alpha(t))]$ consists of the bare matter Hamiltonian and the laser-field driving term.
When the optical frequency is $\omega$, then $\alpha(t)=\alpha \exp(-\mathrm{i} \omega t)$.
The subscripts $\mathrm{m}$, $\mathrm{p}$, and $\mathrm{mp}$ denote the Hilbert spaces of matter, photons, and the total system, respectively (omitted when unambiguous).

The XFA becomes asymptotically valid in the thermodynamic limit for the photon field.
In this limit, the field strength is kept finite while the coupling constant tends to zero and the photon number tends to infinity:
\begin{equation}
g \sqrt{\langle \hat{n} \rangle} = \mathrm{const},\ g \to 0. \label{eq:def:photoninc_thermodynamic_limit}
\end{equation}
As a practical criterion, the XFA works well when the initial photon number $\langle \hat{n} \rangle$ greatly exceeds the number of excited particles $N_{\mathrm{exc}}$, namely, $\langle \hat{n} \rangle \gg N_{\mathrm{exc}}$.
In addition, because this approximation is based on an iterative expansion of the Born series, the condition $g t \ll 1$ also provides a useful estimate of its timescale of validity.
The coherent-state overcompleteness and the resulting nonuniqueness of $\chi(\alpha)$ [Eq.~\eqref{eq:def:initial_photon_state}] do not affect the leading-order XFA dynamics in this limit, since coherent states with different field strengths become orthogonal:
$|\langle F/g| F'/g\rangle|
=\exp[-|F-F'|^2/(2g^2)]
\to 2\pi g^2\,\delta^{(2)}(F-F')$.
Physically, the intrinsic $\mathcal{O}(1)$ quantum uncertainty of a coherent state in $\alpha$ is reduced to a vanishing field uncertainty $\Delta F \sim g$, whereas $\mathcal{O}(1)$ structures in $F$, such as macroscopic quantum superpositions and squeezing, remain.
In summary, the XFA provides a quantum generalization of the external-field picture: it adheres to the superposition principle while neglecting matter-to-light backaction along each laser-driven branch.

\subsection{Squeezed vacuum state and its Janszky representation} \label{sec:preliminaries:Janszky}
Here we adopt the Janszky representation~\cite{Janszky1990} as the coherent-state expansion~\eqref{eq:def:initial_photon_state} of the SV state.
A useful feature of the Janszky representation is that it exploits the overcompleteness of coherent states to simplify an expansion that is, in general, given by a two-dimensional integral over the complex plane into a one-dimensional real integral.

The SV state is defined by
\begin{equation}
|\xi\rangle^{\mathrm{sv}} = \mathrm{e}^{\frac{1}{2}(\xi^{*} \hat{a}^2 - \xi \hat{a}^{\dagger}{}^2)}|0\rangle, \label{eq:def:sv}
\end{equation}
where $\xi = r \exp(\mathrm{i} \vartheta)$ is the squeezing parameter.
Below, we set the angle $\vartheta = 0$ and consider real $r$ only.

The Janszky representation of the SV state is given by~\cite{Janszky1990}
\begin{equation}
|r \rangle^{\mathrm{sv}} = \int_{-\infty}^{\infty} \frac{\mathrm{d}p}{\sqrt{2\pi \sinh r}}\, \mathrm{e}^{-\frac{\coth r -1}{2} p^2} |\mathrm{i}p\rangle. \label{eq:def:sv_Janszky}
\end{equation}
This expression shows that the SV state can be written as a Gaussian-weighted quantum superposition of coherent states, with a width determined by the squeezing parameter $r$.
Furthermore, although the quantumness of an SV state is typically associated with squeezing-induced noise reduction, the Janszky representation highlights another aspect of a pure SV state: the superposition along the anti-squeezed direction indicates that anti-squeezing is not merely excess noise, but rather a manifestation of macroscopic quantum superposition.

\section{General theory for BSV-induced matter dynamics} \label{sec:general_theory}
Based on the formulation above, we derive the XFA effective theory for BSV-light driving.
The key control parameter in this theory is the effective field strength of the SV light,
\begin{equation}
F_{\mathrm{c}} \equiv g \mathrm{e}^{r}. \label{eq:def:field_strength_for_BSV}
\end{equation}
The corresponding photonic thermodynamic limit~\eqref{eq:def:photoninc_thermodynamic_limit} becomes
\begin{equation}
F_{\mathrm{c}} = \mathrm{const},\ g \to 0. \label{eq:photonic_thermodynamic_limit_for_BSV}
\end{equation}
In this limit, the SV Janszky representation can be rewritten, with $F=g p$, as
\begin{equation}
|r\rangle^{\mathrm{sv}} \overset{g \to 0}{=} \int_{-\infty}^{\infty} \frac{\mathrm{d}F}{\sqrt{\pi g F_{\mathrm{c}}}}\, \mathrm{e}^{-(F/F_{\mathrm{c}})^2} |\mathrm{i}F/g\rangle. \label{eq:sv_light_XFA}
\end{equation}
That is, the SV state is represented as a superposition of coherent states weighted by a Gaussian distribution with width $F_{\mathrm{c}}$.
At the same time, the total-system wave function becomes
\begin{equation}
|\Psi_{\mathrm{XFA}}(t)\rangle_{\mathrm{mp}} = \int_{-\infty}^{\infty} \frac{\mathrm{d}F}{\sqrt{\pi g F_{\mathrm{c}}}}\,  \mathrm{e}^{-(F/F_{\mathrm{c}})^2}|\psi_{\mathrm{i}F}(t)\rangle_{\mathrm{m}} |\mathrm{i}F(t)/g\rangle_{\mathrm{p}}. \label{eq:XFA_total-system_state}
\end{equation}
This equation indicates that light--matter entanglement is formed through the quantum superposition of light and matter states labeled by the field strength $F$.

\subsection{Unconditional dynamics} \label{sec:general_theory:unconditional}
We first consider unconditional matter dynamics without any measurement on the light.
The matter state is described by the reduced density matrix obtained by tracing out the photonic degrees of freedom:
\begin{equation}
\hat{\rho}_{\mathrm{m}}(t) = \mathrm{Tr}_{\mathrm{p}} [|\Psi(t)\rangle_{\mathrm{mp}} \, {}_{\mathrm{mp}}\langle \Psi(t)|]. \label{eq:def:unconditional}
\end{equation}

By using the orthogonality of coherent states in the photonic thermodynamic limit~\eqref{eq:def:photoninc_thermodynamic_limit},
$\langle \mathrm{i}F/g|\mathrm{i}F'/g\rangle \overset{g \to 0}{=} \sqrt{2\pi}g \delta(F-F')$,
the unconditional matter density matrix follows from Eq.~\eqref{eq:XFA_total-system_state} as
\begin{equation}
\hat{\rho}_{\mathrm{m}} =  \int \frac{\mathrm{d}F}{\sqrt{\pi /2}F_{\mathrm{c}}}\, \mathrm{e}^{-2(F/F_{\mathrm{c}})^2}\, |\psi_{\mathrm{i}F}\rangle \langle\psi_{\mathrm{i}F}|. \label{eq:uncoditional}
\end{equation}
Thus, the matter density matrix is a Gaussian-weighted classical mixture over the $F$-diagonal states $|\psi_{\mathrm{i}F}\rangle \langle\psi_{\mathrm{i}F}|$.
We call such a state classically reproducible, in the sense that the same matter density matrix can be obtained by driving the matter with a classical ensemble of laser fields whose amplitudes are drawn from the positive Gaussian distribution $\exp[-2(F/F_{\mathrm{c}})^2]$.
Thus, although the input SV state is a quantum state of light, the unconditional matter state contains no coherence between different laser-driven branches within the XFA.
If no intrinsic interparticle interaction is present, classical-light driving does not increase the matter-side quantumness (such as entanglement) contained in each $|\psi_{\mathrm{i}F}\rangle$.
In addition, convex quantum measures do not increase under classical mixtures with positive weights.
It therefore follows that no nonclassical phenomenon emerges in unconditional matter dynamics driven solely by SV light.
This conclusion is consistent with Refs.~\cite{Imai2025b, Imai2026}, which studied electronically driven systems under Schr\"{o}dinger cat-state light, and with Ref.~\cite{Gothelf2026}, which discussed the limitations of approximations based on the Husimi function.

\subsection{Heralded dynamics} \label{sec:general_theory:heralded}
Next, we perform a measurement on the light and derive the matter state conditioned on the outcome.
The measurement is described by Kraus operators $\hat{M}_{\mu}$ labeled by outcome $\mu$, and the corresponding effects $\hat{E}_{\mu} = \hat{M}_{\mu}^{\dagger} \hat{M}_{\mu}$ satisfy $\sum_{\mu} \hat{E}_{\mu} = \mathbb{1}$~\cite{Kraus1983, Wiseman2009}.
Using these operators, the heralded matter density matrix at time $t$ is given by
\begin{equation}
\hat{\rho}_{\mathrm{m}}^{\mu}(t) = 
\frac{\mathrm{Tr}_{\mathrm{p}} \left[ \hat{M}_{\mu}|\Psi(t)\rangle_{\mathrm{mp}}\, {}_{\mathrm{mp}}\langle\Psi(t)|\hat{M}_{\mu}^{\dagger} \right] }{ {}_{\mathrm{mp}}\langle \Psi(t)| \hat{E}_{\mu}|\Psi(t)\rangle_{\mathrm{mp}} }.  \label{eq:postselected_state:general}
\end{equation}

Performing a suitable measurement on the light can disentangle the light--matter system while preserving the matter's quantum superposition.
Here we consider measurements of an optical quadrature.
The quadrature operator with axis angle $\varphi$ is defined as
\begin{equation}
\hat{q}_{\varphi} = \left[ \mathrm{e}^{-\mathrm{i}\varphi} \hat{a} + \mathrm{e}^{\mathrm{i}\varphi} \hat{a}^{\dagger} \right]/\sqrt{2}. \label{eq:def:quadrature_operator}
\end{equation}
Let $|q;\varphi\rangle$ denote the eigenstate with eigenvalue $q \in \mathbb{R}$.
Its overlap with a coherent state $|\alpha\rangle$ is
$\langle q;\varphi| \alpha\rangle = \pi^{-1/4} \, \mathrm{e}^{-\mathrm{i}q_0 p_0 + \mathrm{i} \sqrt{2} p_0 q} \, \mathrm{e}^{-(q-\sqrt{2}q_0)^2/2}$,
where $\alpha \exp(-\mathrm{i}\varphi)= q_0 + \mathrm{i} p_0$.

An ideal projective measurement of the quadrature is represented by the Kraus operator
\begin{equation}
\hat{M}_{q}^{\varphi} = |0\rangle \langle q;\varphi|. \label{eq:ideal_quadrature_projector}
\end{equation}
Choosing the quadrature angle in the rotating frame of the harmonic oscillator ($\varphi = -\omega t$) and introducing the scale-invariant outcome $\tilde{q} = q/g$, Eq.~\eqref{eq:XFA_total-system_state} shows that the heralded matter state can be described using the following unnormalized vector:
\begin{equation}
|\psi^{q}(t)\rangle_{\mathrm{m}} \equiv {}_{\mathrm{p}}\langle q;\varphi|\Psi(t)\rangle_{\mathrm{mp}}. \label{eq:def:q-heralded_state_vector}
\end{equation}
The success probability density for realizing this $q$-heralded state is then
\begin{equation}
P(q) = \langle \psi^q|\psi^q \rangle. \label{eq:def:probability_density}
\end{equation}

Let $|\psi_{\mathrm{i}F}(t)\rangle_{\mathrm{m}} = \hat{U}_{F}(t) |\psi(0)\rangle_{\mathrm{m}}$ denote the time evolution of the laser-driven state with field strength~$F$.
We then define the $q$-heralded time-evolution operator as follows:
\begin{equation}
|\psi^{q}(t)\rangle_{\mathrm{m}} \equiv \hat{\mathcal{U}}^q(t) |\psi(0)\rangle_{\mathrm{m}},\ \ \ \hat{\mathcal{U}}^q(t) \overset{\tilde{q}=\mathrm{const.}}{=} \int \frac{\mathrm{d}F}{\pi^{3/4}\sqrt{g F_{\mathrm{c}}}}\, \mathrm{e}^{-(F/F_{\mathrm{c}})^2}\, \mathrm{e}^{\mathrm{i}F\sqrt{2}\tilde{q}} \, \hat{U}_{F}(t). \label{eq:q-heralded_time_evolution}
\end{equation}
Unlike the classically reproducible unconditional case~\eqref{eq:uncoditional}, this effective time-evolution operator shows that the $q$-heralded state $|\psi^q(t)\rangle_{\mathrm{m}}$ is a Gaussian-weighted quantum superposition of laser-driven states $|\psi_{\mathrm{i}F}(t)\rangle_{\mathrm{m}}$.
The nature of the resulting matter state depends on the details of the laser-driven evolution $\hat{U}_{F}(t)$.
In Sec.~\ref{sec:tavis-cummings} we show that this yields a Dicke state in the TC model, while Sec.~\ref{sec:dicke} demonstrates a stroboscopic cat-like state in the Dicke model.

\section{Application to the Tavis--Cummings model} \label{sec:tavis-cummings}
In this section we analyze the TC model to study concrete matter dynamics.
The model describes the interaction between a resonant single-mode photon field and an ensemble of $N$ independent particles treated as electric-dipole-coupled two-level electronic states (including atoms, excitons, and electron--hole pairs) within the RWA.
After introducing the model and the observables used in the analysis in Sec.~\ref{sec:tavis-cummings:model}, we study quadrature-heralded matter dynamics in Sec.~\ref{sec:tavis-cummings:heralded}, measurement resolution and success probability in Sec.~\ref{sec:tavis-cummings:resolution}, and the validity of the XFA in Sec.~\ref{sec:tavis-cummings:full}.
Unless stated otherwise, the dynamics below is analyzed in the rotating frame at frequency $\omega$.

\subsection{Model and observables} \label{sec:tavis-cummings:model}
We define the $N$-particle TC model~\cite{Tavis1968} as
\begin{equation}
\hat{H}_{\mathrm{TC}}= \omega \hat{a}^{\dagger} \hat{a} + \sum_{j=1}^{N} \varDelta \hat{S}^z_j - \mathrm{i}g \sum_{j} \left( \hat{a} \hat{S}^+_j - \hat{a}^{\dagger} \hat{S}^-_j \right). \label{eq:def:tc_hamiltonian}
\end{equation}
Here, $\hat{S}_j^{b} = \sigma_j^b/2$ ($b=x,y,z$) is the spin-$1/2$ operator representing the $j$th electric-dipole-coupled two-level state, and $\sigma^b$ is a Pauli matrix ($\hat{S}^{\pm}_j=\hat{S}^x_j \pm \mathrm{i} \hat{S}^y_j$).
The parameter $\varDelta$ is the level spacing, and $g$ is the electron--photon coupling constant.
The first and second terms are the free Hamiltonians of the photon field and matter system, respectively, and the third term is the RWA interaction between the electric field and the polarization.
We consider the resonance condition $\omega = \varDelta$, use $\omega$ as the energy unit, and write $\omega=1$ unless otherwise noted.
We also restrict attention to even $N$.

The laser-driven matter state $|\psi_{\mathrm{i}F}(t)\rangle_{\mathrm{m}}$ evolves according to
\begin{equation}
\hat{H}_{\mathrm{TC},\mathrm{m}}[\mathrm{i}F] = \sum_j \varDelta \hat{S}^z_j + \sum_j \left( F(t) \hat{S}^+_j + F(t)^{*} \hat{S}^-_j \right), \label{eq:tc_hamiltonian_electronic_part}
\end{equation}
which is obtained by replacing $\hat{a} \to \mathrm{i}(F/g)$ in Eq.~\eqref{eq:def:tc_hamiltonian}.
Here, $F(t) = F \exp(-\mathrm{i}\omega t)$.
In the $\omega$-rotating frame, the corresponding time-evolution operator is
\begin{equation}
\hat{U}_{\mathrm{TC},F}(t) = \mathrm{e}^{-\mathrm{i}2F\hat{J}^x t}, \label{eq:time_evolution_operator_tc}
\end{equation}
where $\hat{\bm{J}} = \sum_{i} \hat{\bm{S}}_i$ is the collective spin operator.

We take the initial state to be the direct product of the matter ground state, the all-spin-down state $|{\downarrow \downarrow \cdots \downarrow}\rangle$, and the SV state of light $|r\rangle^{\mathrm{sv}}$.
Since the TC model conserves the total spin, it is convenient to use the Dicke basis.
For total spin $J = N/2$, the $b$-axis Dicke states $\{ |J,m\rangle^{b} \mid m = -J,\ldots,J\}$ satisfy
$\hat{\bm{J}}^2|J,m\rangle^{b} = J(J+1)|J,m\rangle^{b}$ and
$\hat{J}^b|J,m\rangle^{b} = m|J,m\rangle^{b}$ for $b=x,y,z$.

Although many measures of quantumness in many-body quantum systems are proposed~\cite{Frowis2018}, we employ the quantum Fisher information (QFI)~\cite{Helstrom1969, Holevo2011, Braunstein1994}, which is well suited for quantifying the number of entangled particles, and the spin Wigner function~\cite{Stratonovich1956, Brif1999}, which is convenient for visualizing quantum states.
The QFI is defined as
\begin{equation}
\mathcal{F}_{\mathrm{Q}} = \underset{\hat{A}}{\mathrm{max}}\, \sum_{i,j} 2\frac{(\lambda_i - \lambda_j)^2}{\lambda_i + \lambda_j} |\langle \lambda_i|\hat{A}|\lambda_j \rangle|^2. \label{eq:def:QFI}
\end{equation}
Here, $\hat{A}$ is an arbitrary additive observable, and $\lambda_i$ and $|\lambda_i\rangle$ are the $i$th eigenvalue and eigenvector of a density matrix $\hat{\rho}$.
For a pure state, $\mathcal{F}_{\mathrm{Q}} = \max_{\hat{A}}[4(\langle \hat{A}^2\rangle - \langle\hat{A}\rangle^2)]$ holds.
An important feature of the QFI is that if the QFI density $\mathcal{F}_{\mathrm{Q}}/N$ exceeds a positive integer $k$, then one can infer that at least $k +1$ particles are entangled~\cite{Toth2012, Hyllus2012, Gessner2019, Ren2021c}.
Thus macroscopic quantumness is present when the QFI density is extensive, namely, $\mathcal{F}_{\mathrm{Q}}/N \propto N$.

Following Ref.~\cite{Davis2021}, we define the spin Wigner function as
\begin{equation}
W(\theta,\phi)=\mathrm{Tr}[\hat{\rho} \, \hat{\Delta}(\theta,\phi)],\ \hat{\Delta}(\theta,\phi)= \sum_{m=-J}^{J} \left( \sum_{j=0}^{2J} \frac{2j+1}{2J+1} C^{Jm}_{Jm;j0}\right) \hat{R}(\theta,\phi)|J,m\rangle^z\, {}^z\langle J,m|\hat{R}^{\dagger}(\theta,\phi), \label{eq:def:spin_Wigner_function}
\end{equation}
where $\hat{R}(\theta,\phi)$ is the rotation operator about the axis $\bm{l} = (-\sin\phi,\cos\phi,0)$, given by $\hat{R}(\theta,\phi) = \mathrm{e}^{\mathrm{i}\phi \hat{J}^z} \mathrm{e}^{\mathrm{i}\theta \hat{J}^y} = \mathrm{e}^{-\mathrm{i}\theta \bm{l} \cdot \hat{\bm{J}}}$, and $C^{JM}_{j_1,m_1;j_2,m_2} = \langle j_1,m_1;j_2,m_2|J,M\rangle$ denotes a Clebsch--Gordan coefficient.

\subsection{Quadrature-heralded dynamics} \label{sec:tavis-cummings:heralded}
We derive the heralded time-evolution operator $\hat{\mathcal{U}}_{\mathrm{TC}}^{q}(t)$ for the matter system conditioned on a quadrature outcome $q$.
From the laser-driven time-evolution operator $\hat{U}_{\mathrm{TC},F}(t)$ [Eq.~\eqref{eq:time_evolution_operator_tc}] and Eq.~\eqref{eq:q-heralded_time_evolution}, we obtain
\begin{equation}
\hat{\mathcal{U}}_{\mathrm{TC}}^{q}(t)  = \sqrt{\frac{F_{\mathrm{c}}}{\sqrt{\pi}g}}\, \mathrm{e}^{-(F_{\mathrm{c}}t)^2 [\hat{J}^x - \tilde{q}/(\sqrt{2}t)]^2}. \label{eq:time_evolution_XFA}
\end{equation}
This shows that SV-light driving followed by a single-shot quadrature projective measurement of the light yields an effective Gaussian filter with respect to $\hat{J}^x$, or equivalently, imaginary-time one-axis twisting~\cite{Kitagawa1993}, for the matter system.
For $\tilde{q} = 0$, this filter selects the $\hat{J}^x = 0$ eigenspace of the collective electric-polarization operator, thereby driving the state toward the $x$-axis Dicke state $|J,0\rangle^x$.
Returning to the laboratory frame, this corresponds to the zero-eigenvalue state of $\cos(\omega t) \hat{J}^x + \sin(\omega t) \hat{J}^y$, where $\hat{J}^y$ is the polarization-current operator.
Moreover, because the asymptotic rate increases with the effective field strength $F_{\mathrm{c}}$ [Eq.~\eqref{eq:def:field_strength_for_BSV}], BSV light enables ultrafast Dicke-state preparation.
Here, ``ultrafast'' means that the preparation time is set by $F_{\mathrm{c}}^{-1}$, which remains finite even in the photonic thermodynamic limit~\eqref{eq:def:photoninc_thermodynamic_limit} and can be shorter than the relaxation and dephasing times of the matter system by using sufficiently bright light.
This contrasts with QND-based Gaussian filters~\cite{Zhang2019h, Vanderbruggen2011, Vanderbruggen2014, Bouchoule2002} and dissipative imaginary-time one-axis twisting~\cite{Groiseau2021, Groiseau2025}, whose timescales are instead set by measurement or dissipative rates.
Even for outcomes $\tilde{q} \neq 0$, the dynamics likewise converges to the $\hat{J}^x = 0$ eigenstate in the long-time regime $t \gg |\tilde{q}|$.

We now analyze the QFI dynamics starting from the initial matter state $|J,-J\rangle^z$.
In the rotating frame, the additive observable that maximizes the QFI~\eqref{eq:def:QFI} is $\hat{A}=\hat{J}^y$.
By expanding the quadrature-heralded state vector $|\psi^q(t)\rangle$ [Eq.~\eqref{eq:q-heralded_time_evolution}] in the $x$-axis Dicke basis, we obtain
\begin{equation}
\mathcal{F}_{\mathrm{Q}} = N+2(1-\mathrm{e}^{-2\tau^2})\frac{\sum_{m=-J}^{J} (J^2 - m^2) |c_m|^2}{\sum_{m=-J}^{J} |c_m|^2}, \label{eq:QFI_XFA_exact}
\end{equation}
where $|\psi^q(t)\rangle = \sum_m c_m |J,m\rangle^x$ with
$c_m = 2^{-J}\sqrt{\binom{2J}{J+m}} \exp[-\tau^2 (m-\mu)^2]$,
$\tau = F_{\mathrm{c}}t$, and $\mu = \tilde{q}/(\sqrt{2}t)$.
For the outcome $\tilde{q} = 0$, this simplifies in the large-$N$ limit to
\begin{equation}
\mathcal{F}_{\mathrm{Q}} \approx N + (1-\mathrm{e}^{-2\tau^2}) \left( \frac{N^2}{2} - \frac{N}{2(1+N\tau^2)} \right). \label{eq:QFI_XFA_large-N}
\end{equation}
This expression demonstrates that the QFI density approaches the saturation value $N/2 + 1$ in a squared-exponential manner, on the timescale of order $F_{\mathrm{c}}^{-1}$ [Eq.~\eqref{eq:def:field_strength_for_BSV}].

\begin{figure}[t]
\centering
\includegraphics[width=\columnwidth]{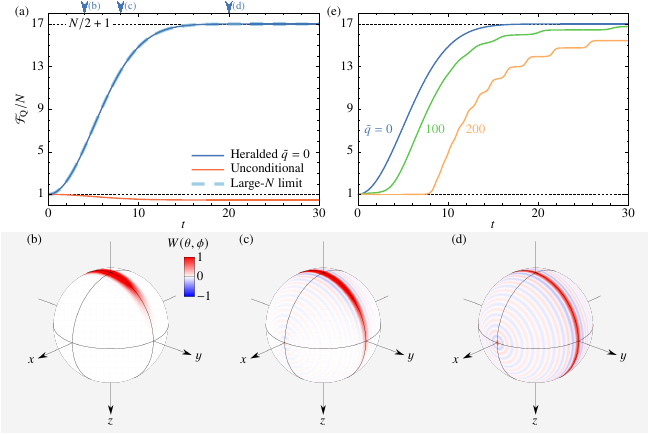}
\caption{
Quadrature-heralded SV-driven matter dynamics in the TC model~\eqref{eq:def:tc_hamiltonian} within the XFA for $F_{\mathrm{c}} \approx 0.10$ ($g=0.005$, $r=3$) and $N = 32$.
(a)~Time profile of the QFI density $\mathcal{F}_{\mathrm{Q}}/N$ (blue solid line), for $\tilde{q} = 0$ [Eq.~\eqref{eq:QFI_XFA_exact}].
The orange line shows the unconditional SV-driven case calculated from $\hat{\rho}_{\mathrm{m}}$ [Eq.~\eqref{eq:uncoditional}].
The light-blue dashed line shows the large-$N$ limit~\eqref{eq:QFI_XFA_large-N}.
(b)--(d)~Spin Wigner function $W(\theta,\phi)$ [Eq.~\eqref{eq:def:spin_Wigner_function}] at the times indicated in (a).
For visualization, only the $z$ axis is inverted.
(e)~QFI dynamics for different quadrature outcomes, $\tilde{q}=q/g=0,\,100,\,200$, corresponding to $q=0,\,0.5,\,1$ for $g=0.005$.
}
\label{fig:qideals}
\end{figure}
Figure~\ref{fig:qideals} shows the numerically calculated quadrature-heralded matter dynamics for $N = 32$ and $F_{\mathrm{c}}=g\exp(r) \approx 0.10$ ($g=0.005$, $r=3$).
Figure~\ref{fig:qideals}(a) plots the QFI density $\mathcal{F}_{\mathrm{Q}}/N$ conditioned on $\tilde{q} = 0$ [Eq.~\eqref{eq:QFI_XFA_exact}] as a blue line, together with the corresponding result calculated from the unconditional SV-driven state $\hat{\rho}_{\mathrm{m}}$ [Eq.~\eqref{eq:uncoditional}] and Eq.~\eqref{eq:def:QFI} as an orange line.
Numerically, the integral over $F$ is evaluated using the rectangular method.
For the unconditional state, the QFI density remains below unity; consequently, the QFI does not reveal any many-body quantumness.
By contrast, with quadrature heralding, the QFI density increases rapidly on a timescale $t \sim F_{\mathrm{c}}^{-1} \approx 10$ and reaches the maximum value $N/2+1$.
This value coincides with the QFI density of the Dicke state $|J,0\rangle^x$.
For comparison, the large-$N$ approximation~\eqref{eq:QFI_XFA_large-N} is also shown as a light-blue dashed line, indicating that even at $N=32$ it already agrees well with the exact result within the XFA~\eqref{eq:QFI_XFA_exact}.

Figures~\ref{fig:qideals}(b)--(d) display spherical color maps of the spin Wigner function $W(\theta,\phi)$ [Eq.~\eqref{eq:def:spin_Wigner_function}], calculated from the normalized state vector $|\psi^q\rangle/\sqrt{\langle\psi^q|\psi^q\rangle}$ [Eq.~\eqref{eq:q-heralded_time_evolution}] at the times marked by the arrowheads above Fig.~\ref{fig:qideals}(a).
In the initial stage of the QFI growth [Fig.~\ref{fig:qideals}(b)], the Gaussian distribution, which is localized around $-z$ at $t=0$, becomes anti-squeezed along the $y$ axis and squeezed along the conjugate $x$ axis.
Near the end of the QFI growth [Fig.~\ref{fig:qideals}(c)], as the squeezing and anti-squeezing develop, negative interference fringes begin to appear around the distribution.
Once the QFI reaches its stationary value [Fig.~\ref{fig:qideals}(d)], the distribution becomes rotationally symmetric about the $x$ axis and is concentrated around the $yz$ plane, with distinct interference fringes, clearly indicating the formation of the Dicke state $|J,0\rangle^x$.
This non-Gaussianity originates from the compact spin phase space inherent to the finite-size matter system.
Under BSV light driving---equivalent to coherent-state driving with a broad amplitude distribution [Eq.~\eqref{eq:XFA_total-system_state}]---each coherent-state branch induces a Rabi oscillation with an amplitude-dependent angular velocity in the TC model.
The resulting trajectories spread along the $y$ axis and wrap around the compact Bloch sphere, yielding a non-Gaussian, $x$-axis rotationally symmetric distribution.
Optical-quadrature heralding then coherently recombines these amplitude-dependent trajectories to extract their latent quantum coherence, generating the negative interference fringes.

Figure~\ref{fig:qideals}(e) shows the dependence of the QFI dynamics on the quadrature measurement outcome.
The $\tilde{q} (=q/g) = 0$ curve (blue) exhibits a squared-exponential increase [see Eq.~\eqref{eq:QFI_XFA_large-N}], whereas the $\tilde{q} = 100$ (green) and $\tilde{q} = 200$ (orange) curves show qualitatively similar behavior, except for a delayed onset and step-like increments at later times.
The later-time structure visible for $t \gtrsim 10$ reflects the discrete spectrum of $\hat{J}^x$.
Importantly, even for outcomes with $\tilde{q} \neq 0$, the heralded state eventually converges to the same state $|J,0\rangle^x$, indicating a certain robustness against finite measurement resolution (Sec.~\ref{sec:tavis-cummings:resolution}).

\subsection{Effects of measurement resolution and success probability} \label{sec:tavis-cummings:resolution}
In this section we investigate the effects of measurement resolution and success probability on the generation of the Dicke state by SV-light driving and quadrature-based heralding.
In this paper, a quadrature measurement with a finite bin width $\delta q_{\mathrm{bin}}$ is modeled by the following effect:
\begin{equation}
\hat{E}_{q}^{\varphi;\delta q_{\mathrm{bin}}} =  \int_{q-\delta q_{\mathrm{bin}}/2}^{q+\delta q_{\mathrm{bin}}/2} \mathrm{d}q'\, |q';\varphi\rangle \langle q';\varphi|. \label{eq:finite_quadrature_projector}
\end{equation}
Using this operator and the measurement probability density $P(q)$ [Eq.~\eqref{eq:def:probability_density}], the matter state heralded by an outcome centered at $q$ with resolution $\delta q_{\mathrm{bin}}$ is given by
\begin{equation}
\hat{\rho}^{q;\delta q_{\mathrm{bin}}} = \frac{\int_{q-\delta q_{\mathrm{bin}}/2}^{q+\delta q_{\mathrm{bin}}/2} \mathrm{d}q' |\psi^{q'}\rangle \langle \psi^{q'}|}{\int_{q-\delta q_{\mathrm{bin}}/2}^{q+\delta q_{\mathrm{bin}}/2} \mathrm{d}q' P(q')}. \label{eq:finite_resolved_matter_density_matrix}
\end{equation}

\begin{figure}[t]
\centering
\includegraphics[width=0.6\columnwidth]{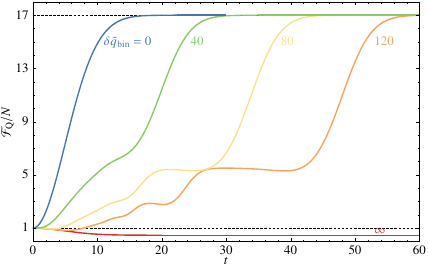}
\caption{
Dependence of the quadrature-heralded QFI density $\mathcal{F}_{\mathrm{Q}}/N$ on the measurement resolution $\delta \tilde{q}_{\mathrm{bin}}$ in the TC model~\eqref{eq:def:tc_hamiltonian} within the XFA for $F_{\mathrm{c}} \approx 0.10$ ($g = 0.005$, $r = 3$), $\tilde{q} = 0$, and $N = 32$.
The QFI is calculated from $\hat{\rho}^{q;\delta q_{\mathrm{bin}}}$ [Eq.~\eqref{eq:finite_resolved_matter_density_matrix}].
Given $g=0.005$, the resolutions $\delta \tilde{q}_{\mathrm{bin}} = \delta q_{\mathrm{bin}}/g = 40,\,80,\,120$ correspond to $\delta q_{\mathrm{bin}} = 0.2,\,0.4,\,0.6$, respectively.
}
\label{fig:qcoarse-grain}
\end{figure}
Figure~\ref{fig:qcoarse-grain} shows the time evolution of the QFI density $\mathcal{F}_{\mathrm{Q}}/N$ [Eq.~\eqref{eq:def:QFI}] calculated from the matter state $\hat{\rho}^{q;\delta q_{\mathrm{bin}}}$ [Eq.~\eqref{eq:finite_resolved_matter_density_matrix}] heralded by a finite-resolution quadrature measurement.
Since the scaled outcome $\tilde{q}=q/g$ is the relevant quantity in the XFA~\eqref{eq:q-heralded_time_evolution}, the dynamics depends on $\delta q_{\mathrm{bin}}$ only through the scaled resolution $\delta \tilde{q}_{\mathrm{bin}} = \delta q_{\mathrm{bin}}/g$.
We set the outcome center to $\tilde{q} = 0$.
Numerically, the integral over $q'$ is evaluated using the rectangular method.
Overall, the monotonic growth of the ideal measurement $\delta \tilde{q}_{\mathrm{bin}} = 0$ is suppressed up to a certain time by finite $\delta \tilde{q}_{\mathrm{bin}}$, but eventually crosses over to the same squared-exponential growth.
At early times, larger $\delta \tilde{q}_{\mathrm{bin}}$ prolongs the agreement with the unconditional curve (red), corresponding to $\delta \tilde{q}_{\mathrm{bin}} \to \infty$.
This is followed by stepwise increases arising from the discreteness of $\hat{J}^x$, as in the case of nonzero outcomes $\tilde{q} \neq 0$.
Finally, as $\delta \tilde{q}_{\mathrm{bin}}$ increases, the waiting time before the onset of squared-exponential growth becomes longer.
Below we analyze this behavior separately in the high-resolution limit of small $\delta \tilde{q}_{\mathrm{bin}}$ and the coarse-grained limit of large $\delta \tilde{q}_{\mathrm{bin}}$.

\subsubsection*{High-resolution limit}
We analyze the high-resolution case $\delta q_{\mathrm{bin}} \ll 1$ and examine the particle-number $N$ scaling of the QFI weighted by the success probability.
This quantity is important because it directly determines the practically achievable sensitivity in quantum metrology.

For small resolution $\delta q_{\mathrm{bin}} \ll 1$, the probability-weighted QFI can be approximated as
\begin{equation}
\mathcal{F}_{\mathrm{Q}}[\hat{\rho}^{q;\delta q_{\mathrm{bin}}}] \int_{q-\delta q_{\mathrm{bin}}/2}^{q+\delta q_{\mathrm{bin}}/2}\mathrm{d}q' P(q')  \approx  \mathcal{F}_{\mathrm{Q}}\left[|\psi^{q}\rangle/\sqrt{\langle \psi^{q}|\psi^{q}\rangle} \right]\, P(q)\delta q_{\mathrm{bin}}. \label{eq:def:finite_resolved_QFI}
\end{equation}
For the bin centered at $q = 0$, the probability density $P(0)=\langle \psi^{0}|\psi^{0}\rangle$ reads
\begin{equation}
\langle \psi^{0}|\psi^{0}\rangle = \frac{F_{\mathrm{c}}}{\sqrt{\pi}g}\sum_m 2^{-N}\binom{2J}{J+m} \mathrm{e}^{-2\tau^2 m^2}
\approx
\begin{cases}
\frac{F_{\mathrm{c}}}{\sqrt{\pi}g} \frac{1}{\sqrt{1+N\tau^2}}, & (\tau \ll 1) \\
\frac{F_{\mathrm{c}}}{\sqrt{\pi}g} \frac{1}{\sqrt{\pi N/2}}, & (\tau \gg 1).
\end{cases}
\label{eq:<psi^0|psi^0>}
\end{equation}
In either limit, one thus finds $P(0) \propto N^{-1/2}$.
Meanwhile, Eq.~\eqref{eq:QFI_XFA_large-N} shows that the QFI of the normalized state $|\psi^{q}\rangle$ scales as $\mathcal{F}_{\mathrm{Q}} \propto N^2$ for $\tau = F_{\mathrm{c}} t \gg 1$.
Accordingly, the probability-weighted QFI scales as
\begin{equation}
\mathcal{F}_{\mathrm{Q}}\left[|\psi^{q}\rangle/\sqrt{\langle \psi^{q}|\psi^{q}\rangle} \right]\, P(q)\delta q_{\mathrm{bin}} \propto N^{3/2}. \label{eq:high-resolution:finite_resolved_QFI}
\end{equation}
This result shows that, although the Heisenberg scaling of the Dicke state $|J,0\rangle^x$ is weakened by the success probability, the attainable precision still surpasses the SQL.

We further quantify the success probability.
For the parameters used in Fig.~\ref{fig:qcoarse-grain}, Eq.~\eqref{eq:<psi^0|psi^0>} gives the stationary long-time probability as $P(0) \delta q_{\mathrm{bin}} \simeq 1.6\,\delta q_{\mathrm{bin}}$; for example, this corresponds to approximately $3\%$ for $\delta q_{\mathrm{bin}}=0.02$.
Although this probability is not large, useful heralding events are not restricted to $q=0$.
If quadrature outcomes other than $q=0$ can also be resolved and recorded with sufficient precision, high-QFI Dicke states besides $|J,0\rangle^x$ can be heralded, as shown in Fig.~\ref{fig:qideals}(e).
Indeed, a general Dicke state $|J,m\rangle^x$ satisfies $\mathcal{F}_{\mathrm{Q}}[|J,m\rangle^x]/N = N/2 + 1 - 2m^2/N$.
Thus, Dicke states with $|m| \leq \mathcal{O}(\sqrt{N})$ remain macroscopic quantum states.
This suggests the possibility that including heralding events associated with the Dicke states satisfying $|m| \leq \mathcal{O}(\sqrt{N})$ could compensate for the $N^{-1/2}$ scaling of each individual probability, thereby allowing the probability-weighted conditional QFI to recover Heisenberg scaling.
However, because the finite-time dynamics associated with nonzero quadrature outcomes exhibits delayed onsets and step-like increments [Fig.~\ref{fig:qideals}(e)], a detailed quantitative optimization is left for future work.

Here, we turn to a comparison between driving with SV-state light and driving with even cat-state light, since both the SV state and the even cat state are composed solely of even-photon-number components.
We analyze the metrological scaling of the matter states induced by these drives.
For driving with the even cat-state light $|\alpha_0\rangle + |{-}\alpha_0\rangle$, the total-system wave function in the XFA is given by~\cite{Imai2026}
\begin{equation}
|\Psi_{\mathrm{XFA}}\rangle_{\mathrm{mp}}^{\mathrm{cat}} = \frac{1}{\sqrt{2}} (|\psi_{\alpha_0}\rangle_{\mathrm{m}}|\alpha_0\rangle + |\psi_{-\alpha_0}\rangle_{\mathrm{m}} |{-}\alpha_0\rangle ). \label{eq:def:cat_XFA_total-system_wave_function}
\end{equation}
The probability density for a quadrature outcome $q$ is then
\begin{equation}
P^{\mathrm{cat}}(q) = \frac{\mathrm{e}^{-q^2}}{2\sqrt{\pi}} \left(2+\langle \psi_{\alpha_0}|\psi_{-\alpha_0}\rangle \mathrm{e}^{\mathrm{i}2\sqrt{2}\alpha_0 q} + \langle \psi_{-\alpha_0}|\psi_{\alpha_0}\rangle \mathrm{e}^{-\mathrm{i}2\sqrt{2}\alpha_0 q} \right), \label{eq:cat_probability_density}
\end{equation}
with $\alpha_0 \in \mathbb{R}$ and the quadrature angle $\varphi = -\omega t +\pi/2$.
When the QFI is maximized, namely, $\langle \psi_{-\alpha_0}|\psi_{\alpha_0}\rangle =0$, this reduces to
\begin{equation}
P^{\mathrm{cat}}(q) = \frac{\mathrm{e}^{-q^2}}{\sqrt{\pi}}, \label{eq:cat_probabitily_density_at_QFI_peak}
\end{equation}
which is independent of the particle number $N$.
It follows that
\begin{equation}
\mathcal{F}_{\mathrm{Q}}\left[(|\psi_{\alpha_0}\rangle+|\psi_{-\alpha_0}\rangle)/\sqrt{2} \right]\, P^{\mathrm{cat}}(q)\delta q_{\mathrm{bin}} \propto N^2, \label{eq:high-resolution:finite_resolved_QFI:cat}
\end{equation}
so the Heisenberg limit is retained even after including the success probability.
Thus, while SV light and even cat-state light consist only of even-photon-number states, even cat-state light can generate stronger nonclassicality in the matter system.

\subsubsection*{Coarse-grained limit}
Next, we derive the relationship between the measurement resolution $\delta q_{\mathrm{bin}}$ and the onset time of the squared-exponential growth of the QFI.
At long times, the heralded state can be approximated by a three-component truncation in the $\hat{J}^x$ basis, involving only $m=0,\pm1$, because of the Gaussian filtering $\hat{\mathcal{U}}_{\mathrm{TC}}^q(t) \sim \exp[-(F_{\mathrm{c}}t)^2 (\hat{J}^x)^2]$ [Eq.~\eqref{eq:time_evolution_XFA}].
Then the $q$-heralded state vector follows from Eq.~\eqref{eq:q-heralded_time_evolution} as
\begin{align}
|\psi^q\rangle \approx& 
\sqrt{\frac{F_{\mathrm{c}}}{\sqrt{\pi}g}} 2^{-J} \Biggl( \sqrt{\binom{2J}{J}} \mathrm{e}^{-(F_{\mathrm{c}}\tilde{q})^2/2}|J,0\rangle^x \nonumber \\
&+\sqrt{\binom{2J}{J+1}} \mathrm{e}^{-[F_{\mathrm{c}}(\tilde{q}-\sqrt{2}t)]^2/2}|J,1\rangle^x 
+\sqrt{\binom{2J}{J-1}} \mathrm{e}^{-[F_{\mathrm{c}}(\tilde{q}+\sqrt{2}t)]^2/2}|J,{-}1\rangle^x 
\Biggr). \label{eq:qprojected_state_vector:long-time_approximation}
\end{align}
We estimate the convergence time toward $|J,0\rangle^x$ from the admixture of the neighboring components $|J,\pm1\rangle^x$.
For either of the diagonal components $|J,\pm 1\rangle^x {}^x\langle J,\pm 1|$ contained in the density matrix $\hat{\rho}^{q;\delta q_{\mathrm{bin}}}$ [Eq.~\eqref{eq:finite_resolved_matter_density_matrix}], the coefficient is proportional to
\begin{equation}
\int_{-\delta q_{\mathrm{bin}}/2}^{\delta q_{\mathrm{bin}}/2} \mathrm{d}q' \frac{F_{\mathrm{c}}}{\sqrt{\pi}g} 2^{-N} \binom{2J}{J\pm1} \mathrm{e}^{-F_{\mathrm{c}}^2(q'/g \mp \sqrt{2}t)^2} \approx \frac{1}{2\pi F_{\mathrm{c}}\sqrt{N}} \frac{\mathrm{e}^{-2F_{\mathrm{c}}^2 [t-\delta q_{\mathrm{bin}} /(2\sqrt{2}g)]^2}}{t-\delta q_{\mathrm{bin}}/(2\sqrt{2}g)}, \label{eq:diagonal_coefficients}
\end{equation}
which is an expansion controlled by $\sqrt{2}F_{\mathrm{c}}[t-\delta q_{\mathrm{bin}}/(2\sqrt{2}g)] \gg 1$.
Hence, the crossover time is estimated as
\begin{equation}
t \gtrsim \frac{\delta \tilde{q}_{\mathrm{bin}}}{2\sqrt{2}}, \label{eq:time_resolution}
\end{equation}
beyond which the asymptotic dynamics approaching the Dicke state $|J,0\rangle^x$ appear even for finite-resolution measurements.

Once the state has converged to $|J,0\rangle^x$, the stationary heralding probability is given by $ \int_{-\delta q_{\mathrm{bin}}/2}^{\delta q_{\mathrm{bin}}/2} dq\,\frac{F_\mathrm{c}}{\sqrt{\pi}g}   2^{-N} \allowbreak \binom{N}{N/2}  \mathrm{e}^{-(F_{\mathrm{c}} q/g)^2} = 2^{-N} \binom{N}{N/2}\, \mathrm{erf}(F_{\mathrm{c}}\delta q_{\mathrm{bin}}/(2g)) \to \sqrt{\frac{2}{\pi N}}$ for $\delta \tilde{q}_{\mathrm{bin}} \gg F_{\mathrm{c}}^{-1}$ and $N \gg 1$. 
Thus, the success probability scales as $N^{-1/2}$, as in the high-resolution limit [Eq.~\eqref{eq:<psi^0|psi^0>}], rather than decaying exponentially with $N$.
For $N=32$, as considered in Fig.~\ref{fig:qcoarse-grain}, the asymptotic value is approximately $0.14$.

\subsection{Validity of XFA: comparison with total-system simulations} \label{sec:tavis-cummings:full}
We validate the XFA used above by comparing it with numerical simulations of the total system.
We compute the time evolution of the total-system state under the TC Hamiltonian $\hat{H}_{\mathrm{TC}}$ [Eq.~\eqref{eq:def:tc_hamiltonian}] using a small time step $\updelta t$ as $|\Psi(t+\updelta t)\rangle_{\mathrm{mp}} = \exp(-\mathrm{i}\hat{H} \updelta t)|\Psi(t)\rangle_{\mathrm{mp}} = \sum_{n=0}^{4} (-\mathrm{i}\hat{H} \updelta t)^n/n!\,|\Psi(t)\rangle_{\mathrm{mp}} + \mathcal{O}(\updelta t^5)$.
The numerical accuracy is controlled by $\updelta t$ and by the truncation maximum photon number $n_{\mathrm{max}}$ of the photonic Hilbert space.
Convergence is confirmed by setting $(\updelta t, n_{\mathrm{max}})$ as: $(10^{-4}, 2000)$ for $g=0.02$ and $0.015$; $(10^{-4}, 3000)$ for $g=0.01$; and $(1.25 \times 10^{-5}, 5000)$ for $g=0.005$.

\begin{figure}[t]
\centering
\includegraphics[width=0.6\columnwidth]{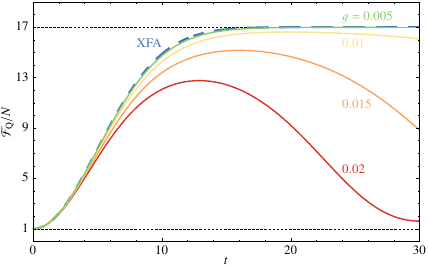}
\caption{
Comparison between the XFA and full light--matter simulations for the quadrature-heralded QFI density $\mathcal{F}_{\mathrm{Q}}/N$ in the TC model~\eqref{eq:def:tc_hamiltonian}.
The blue dashed curve shows the XFA result~\eqref{eq:QFI_XFA_exact}, and the solid curves show full numerical simulations for different electron--photon couplings $g = 0.005,\,0.01,\,0.015,$ and $0.02$.
Parameters are set to $r = 3 - \ln(g/0.005)$, $q = 0$, $\delta q_{\mathrm{bin}} =0$, and $N = 32$.
}
\label{fig:qfull}
\end{figure}
Figure~\ref{fig:qfull} shows the dependence of the quadrature-heralded QFI density $\mathcal{F}_{\mathrm{Q}}/N$ dynamics on the electron--photon coupling constant $g$, where we assume the ideal projection~\eqref{eq:ideal_quadrature_projector} for $q = 0$ and the density matrix~\eqref{eq:postselected_state:general}.
The squeezing parameter is set to $r=3-\ln(g/0.005)$ so as to keep $F_{\mathrm{c}}$ [Eq.~\eqref{eq:def:field_strength_for_BSV}].
As $g$ is reduced from $0.02$ to $0.005$, the full simulation agrees with the XFA (blue dashed curve) over an increasingly long time span.
This observation is consistent with the Born-series-based formulation of the XFA, which is justified when $g t \ll 1$.
The long-time deviation between the exact numerical results and the XFA can be attributed to backaction from the driven matter system onto the quantum light; its physical consequences are summarized in~\ref{sec:extensions:unconditional}.

\section{Application to the Dicke model} \label{sec:dicke}
In this section, we analyze the Dicke model~\cite{Rabi1937, Dicke1954} as a light--matter coupling model without the RWA.
We show that the counter-rotating terms can periodically reshape the heralded Dicke state into a cat-like state.

The $N$-qubit Dicke model is defined as
\begin{align}
&\hat{H}_{\mathrm{D}} = \omega \hat{a}^{\dagger} \hat{a} + \sum_{j=1}^{N} \varDelta \hat{S}^z_j - \hat{E} \cdot \hat{P}, \label{eq:def:Dicke_Hamiltonian} \\
&\hat{E}=\mathrm{i} E_0 (\hat{a} - \hat{a}^{\dagger}),\ \hat{P}=\sum_{j} d (\hat{S}^+_j + \hat{S}^-_j). \label{eq:def:electric_field_polarization}
\end{align}
Here, $\hat{E}$ is the electric-field operator ($E_0$ is the mode function), and $\hat{P}$ is the electric-polarization operator ($d$ is the electric-dipole moment).
The TC model~\eqref{eq:def:tc_hamiltonian} is obtained by applying the RWA, which neglects the counter-rotating terms $\hat{a} \hat{S}^-_j$ and $\hat{a}^{\dagger} \hat{S}^+_j$ oscillating at $2\omega$ in the rotating frame.
In this case, $g=d E_0$.
The RWA is valid when one focuses on the coarse-grained dynamics obtained by averaging over the rapid $2\omega$ oscillations, or when the field strength $g \sqrt{\langle \hat{n} \rangle}$ is not too large.

We analyze the quadrature-heralded matter state $|\psi^q(t)\rangle$ [Eq.~\eqref{eq:q-heralded_time_evolution}] in the Dicke model within the XFA.
The laser-driven matter state $|\psi_{\mathrm{i}F}(t)\rangle$ evolves under the Hamiltonian $\hat{H}_{\mathrm{D,m}}[\mathrm{i}F] = \omega \hat{J}^z + 2(F \mathrm{e}^{-\mathrm{i}\omega t} + F^* \mathrm{e}^{\mathrm{i}\omega t})\hat{J}^x$, obtained by replacing $\hat{a} \to \mathrm{i}(F/g)$.
Numerically, we diagonalize the single-two-level Hamiltonian $\hat{h}_{\mathrm{D,m},j}$ ($\hat{H}_{\mathrm{D,m}}=\sum_j \hat{h}_{\mathrm{D,m},j}$), evolve each two-level state according to $|\psi_j(t+\updelta t)\rangle = \exp(-\mathrm{i} \hat{h}_{\mathrm{D,m},j} \updelta t) |\psi_j(t)\rangle$, and construct the $N$-qubit state as their direct product ($\updelta t = 0.01$).
For the optical quadrature measurement, as in Sec.~\ref{sec:tavis-cummings:heralded}, we consider a projective measurement onto a quadrature eigenstate~\eqref{eq:ideal_quadrature_projector}.

\begin{figure}[t]
\centering
\includegraphics[width=\columnwidth]{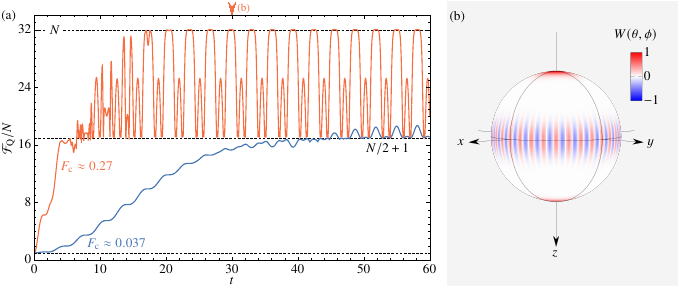}
\caption{
Quadrature-heralded matter dynamics in the Dicke model~\eqref{eq:def:Dicke_Hamiltonian} within the XFA for $\tilde{q} = 0$ and $N = 32$.
(a)~Time profile of the QFI density $\mathcal{F}_{\mathrm{Q}}/N$ for $F_{\mathrm{c}} = g \exp(r) \approx 0.27$ and $0.037$, corresponding to $r = 4$ and $2$ for $g = 0.005$.
(b)~Spin Wigner function $W(\theta,\phi)$ at the time indicated in (a).
}
\label{fig:rds}
\end{figure}
Figure~\ref{fig:rds} shows the quadrature-heralded QFI~\eqref{eq:def:QFI} and spin Wigner function~\eqref{eq:def:spin_Wigner_function}, calculated from the normalized state vector $|\psi^q\rangle/\sqrt{\langle\psi^q|\psi^q\rangle}$ [Eq.~\eqref{eq:q-heralded_time_evolution}] for $\tilde {q}=0$.
In Fig.~\ref{fig:rds}(a), the QFI exhibits an oscillatory component at frequency $2\omega$ ($\omega=1$).
For the relatively weak SV field strength, $F_{\mathrm{c}}=g \exp(r) \approx 0.037$ ($r=2$, $g=0.005$), this oscillation remains small, consistent with the RWA results for the TC model [Eq.~\eqref{eq:QFI_XFA_large-N} and Fig.~\ref{fig:qideals}(a)].
By contrast, for the stronger field $F_{\mathrm{c}}\approx 0.27$ ($r=4$), the oscillation amplitude becomes large, and the QFI density reaches its maximal value $N$ at certain times.
The corresponding spin Wigner functions in Fig.~\ref{fig:rds}(b), taken at the arrow-marked times in Fig.~\ref{fig:rds}(a), display Gaussian lobes localized at the $\pm z$ axes together with clear interference fringes on the $xy$ plane.
These indicate the formation of a $z$-polarized cat-like state, namely a Greenberger--Horne--Zeilinger (GHZ)-like state~\cite{Greenberger1989}.

To understand heuristically the mechanism of this $z$-polarized cat-like state formation, we perform an analytical calculation using the Floquet--Magnus expansion~\cite{Magnus1954, Shirley1965, Sambe1973, Rahav2003, Eckardt2015}.
In this approach, the time-evolution operator $\hat{U}_{\mathrm{D},F}(t)$ for the laser-coupled Dicke model $\hat{H}_{\mathrm{D,m}}[\mathrm{i}F]$ can be approximated as
\begin{align}
&\hat{U}_{\mathrm{D},F}(t) \approx \mathrm{e}^{-\mathrm{i} \hat{K}(t)} \, \hat{U}_{\mathrm{TC},F}(t), \label{eq:def:Floquet-Magnus} \\
&\hat{K}(t) = \frac{F}{\omega} \left[ \sin (2\omega t) \hat{J}^x + (\cos(2\omega t) -1) \hat{J}^y \right]. \label{eq:kick_operator:Dicke}
\end{align}
This approximation decomposes the Dicke-model evolution into the RWA evolution $\hat{U}_{\mathrm{TC},F}(t)$ [Eq.~\eqref{eq:time_evolution_operator_tc}] and a micromotion operator $\exp[-\mathrm{i} \hat{K}(t)]$ oscillating at frequency $2\omega$, and is valid under the condition $F/\omega \ll 1$.

We focus on times $t=(2m+1)\pi/(2\omega)$ ($m \in \mathbb{N}$), at which the noncommutativity between the micromotion operator and the RWA time-evolution operator becomes pronounced.
At these times, the $\tilde{q} = 0$ quadrature-heralded time-evolution operator under the XFA is calculated from Eq.~\eqref{eq:q-heralded_time_evolution} as
\begin{equation}
\hat{\mathcal{U}}_{\mathrm{D}}^{q=0}(t) \propto \sum_{m_y,m_x} \mathrm{e}^{-(F_{\mathrm{c}}t)^2 [m_x - m_y/(\omega t)]^2}\, {}^y\langle J,m_y|J,m_x\rangle^x\,  |J,m_y\rangle^y\, \, {}^x\langle J,m_x|. \label{eq:time_evolution_XFA:Dicke}
\end{equation}
Examining the exponential factor, we find that at sufficiently late times ($t \gg F_{\mathrm{c}}^{-1}$) the main contribution first comes from the collective-spin eigenvalue $m_x = 0$ along the $x$ axis.
This agrees with the RWA result obtained in Sec.~\ref{sec:tavis-cummings:heralded}.
On top of this, the correction factor from the counter-rotating terms is expressed as
\begin{equation}
\hat{\mathcal{U}}_{\mathrm{D}}^{q=0}(t) \propto \mathrm{e}^{-\left(\frac{F_{\mathrm{c}}}{\omega}\right)^2 (\hat{J}^y)^2} |J,0\rangle^x\,{}^x\langle J,0|. \label{eq:time_evolution_XFA:Rabi:y-squeezing}
\end{equation}
The dynamics can therefore be interpreted as a two-step filtering process: Gaussian filtering first selects $|J,0\rangle^x$ in the $x$-spin basis, and the counter-rotating terms then impose an additional Gaussian filter on the $y$ spin with strength $F_{\mathrm{c}}/\omega$.
When the SV field strength $F_{\mathrm{c}}/\omega$ is small, this additional filtering along the $y$ axis has only a minor effect, as seen in the blue curve in Fig.~\ref{fig:rds}(a).
As $F_{\mathrm{c}}/\omega$ increases, however, the spin Wigner distribution, initially spread isotropically over the $yz$ plane, is filtered toward the eigenspace of $\hat{J}^y = 0$, whose weight is concentrated around the $xz$ plane.
As a result, the weight shifts toward the $\pm z$ directions, and a $z$-polarized cat-like state emerges during this intermediate process.

This mechanism admits a simple analytical interpretation, particularly for $F_{\mathrm{c}}/\omega = N^{-1/2}$.
Expanding $|J,0\rangle^x = \sum_m c_m |J,m\rangle^y$, one finds that within the Gaussian filtering $\exp(-m^2/N)$ the coefficients $c_m$ can be regarded as nearly uniform magnitude, up to symmetry-related phases.
That is, $c_m \propto N^{-1/2} \Pi_J(m)$, where $\Pi_J(m)=[1+(-1)^{J-m}]/2$.
Combining this with the large-$N$ expression $|J,\pm J\rangle^z \propto N^{-1/4} \sum_m (\pm 1)^m \allowbreak \exp(-m^2/N) |J,m\rangle^y$, we obtain
\begin{equation}
\mathrm{e}^{-(\hat{J}^y)^2/N} |J,0\rangle^x \propto N^{-1/4} \left(|J,J\rangle^{z} + (-1)^J |J,-J\rangle^z \right). \label{eq:Dicke:z-axis_cat}
\end{equation}
The above argument clarifies that the counter-rotating terms in the Dicke model drive a stroboscopic transformation of the $x$-axis Dicke state $|J,0\rangle^x$ into the $z$-polarized cat-like state.

\section{Conclusion} \label{sec:conclusion}
We propose a framework for the ultrafast generation of macroscopic quantum states in matter using bright squeezed vacuum (BSV) light.
In the weak-coupling, multiphoton regime (the photonic thermodynamic limit), we show that driving with BSV light alone yields only a classical mixture of laser-driven matter states because of light--matter entanglement.
We then demonstrate that, when the system is heralded by the outcome of a single-shot quadrature projective measurement on the post-interaction squeezed-vacuum (SV) light, the resulting matter state generally becomes a Gaussian-weighted quantum superposition.
For a minimal model with electric-field--polarization coupling, the heralded matter system is filtered into the zero-eigenspace of the matter observable coupled to light, at a rate proportional to the square root of the photon number in the SV light.
The resulting macroscopic quantum states also exhibit metrological performance beyond the standard quantum limit.
These results suggest that macroscopic quantum states can be prepared on ultrashort timescales, before relaxation or dephasing becomes dominant, thereby opening a new route to ultrafast quantum metrology.

The proposed mechanism for generating macroscopic quantum states in matter can be extended to a broader class of matter systems, including atoms, molecules, and solid-state electronic systems.
In such systems, the dependence of the laser-driven time-evolution operator $\hat{U}_{F}(t)$ on the field strength $F$ generally becomes more intricate, which is expected to give rise to a richer variety of matter states, as illustrated by the nonlinear effects of the counter-rotating terms in Sec.~\ref{sec:dicke}.
As a simple generalization, whenever the $F$-dependent evolution is generated by a matter observable $\hat{O}$ as $\hat{U}_{F}(t) = \exp(-\mathrm{i} 2F \hat{O}t)$, the quadrature-heralded time-evolution operator is obtained as $\hat{\mathcal{U}}^q(t) \propto \exp[-(F_{\mathrm{c}} t)^2 (\hat{O}-\tilde{q}/(\sqrt{2}t))^2]$.
Thus, heralding essentially acts as a Gaussian filter toward the zero eigenspace of the observable $\hat{O}$ coupled to the optical field $F$.
This observation suggests a general design principle for strong-field quantum-light engineering of many-body quantum states.

This viewpoint also clarifies how coupling inhomogeneity modifies the TC-model result.
If the electron--photon coupling in Eq.~\eqref{eq:def:tc_hamiltonian} is made particle dependent, $g_j = g \eta_j$, the matter operator coupled to the optical field changes from $\hat J^x$ to the coupling-weighted polarization $\hat{P}^x \equiv \sum_j \eta_j \hat{S}^x_j$.
Quadrature heralding then filters the state toward the zero eigenspace of $\hat{P}^x$.
Frequency inhomogeneity has a different effect.
Residual detunings introduce terms of the form $\sum_j\delta_j \hat{S}_j^z$, which do not commute with $\hat{J}^x$ or $\hat{P}^x$, and can therefore cause dephasing.
The present study focuses on the ultrafast regime $|\delta_j| \ll F_{\mathrm{c}}$, where the macroscopic quantum state is generated before this dephasing becomes relevant; a detailed analysis of this effect is left for future work.

We summarize experimental parameter estimates below.
In free-space driving, the electron--photon coupling constant $g$ depends on the effective mode volume $V_{\mathrm{eff}}$.
Nevertheless, the standard single-mode expression $g=d\sqrt{\omega/2\hbar \epsilon_0 V_{\mathrm{eff}}}$ gives a typical scale of $g/\omega \sim 10^{-8}$--$10^{-5}$.
In cavity systems, $g/\omega \sim 10^{-2}$--$10^{-1}$ is widely available~\cite{Forn-Diaz2019}.
As for the squeezing parameter $r$, directly measured values reach $r \approx 1.73$ ($15$~dB)~\cite{Vahlbruch2016}.
For recent BSV light, meanwhile, one can infer $r_{\mathrm{eff}} \approx 15$ from photon numbers $\langle \hat{n} \rangle \sim 10^{13}$~\cite{Sh.Iskhakov2012, Chekhova2015a, Sharapova2020}, although this is not a direct measurement.
Assuming $\hbar \omega = 1.55$~eV in the optical/infrared range, $F_{\mathrm{c}}/\omega \gtrsim 0.5$ allows state generation on subfemtosecond, i.e., attosecond, timescales.
This can be achieved, for example, either with $g/\omega \sim 10^{-7}$ and $\langle \hat{n} \rangle \sim 10^{13}$, or with $g/\omega \sim 10^{-1}$ and $r \sim 1.7$.
Meanwhile, the measurement model in Sec.~\ref{sec:tavis-cummings:resolution} indicates that the required quadrature resolution is $\delta q_{\mathrm{bin}} \lesssim 2\sqrt{2} gt$ [Eq.~\eqref{eq:time_resolution}].
To reach the attosecond regime, one likewise needs $\delta q_{\mathrm{bin}} \sim \exp(-r)$.
In particular, for BSV light with $\langle \hat{n}\rangle \sim 10^{13}$, this requires $\delta q_{\mathrm{bin}} \lesssim 10^{-6}$.
Hence, advances in quantum measurement techniques for the multiphoton regime are essential for realizing and observing new attosecond quantum many-body phenomena driven by strong-field quantum light.
Notably, this is also the resolution required to verify the intrinsic quantum nature of the BSV light itself, which is consistent with the idea that resolving the quantum nature of the driving light is a prerequisite for inducing quantum features in the driven system.

It is also useful to compare the success-probability scale with those of other heralded schemes for preparing nonclassical states.
A representative benchmark is generalized photon subtraction~\cite{Takase2021a}, which is used in continuous-variable optical quantum computing.
This all-optical scheme mixes two squeezed states on a beam splitter, and photon-number measurement at one output port heralds a cat state at the other port.
For a realistic parameter set, the success probability was estimated to be about $2.3\%$~\cite{Takase2021a}, which is comparable to the success probabilities estimated in Sec.~\ref{sec:tavis-cummings:resolution}, ranging from a few percent to over ten percent.
This comparison is only a probability benchmark: while fully optical systems benefit from high repetition rates to yield a large number of successful events, further analysis is required to evaluate the actual generation rate in matter-state heralding protocols based on light--matter coupled systems~\cite{Nielsen2009, McConnell2013, McConnell2015, Chen2015b, Masson2019, Nakamoto2026}.

As a future direction, it is important to refine the treatment of the mode structure of the light field.
In this work, the light interacting with matter is modeled as a single bosonic mode.
A more rigorous treatment of scattering, input-output processes, and measurement, however, would require multiple channels corresponding to the input and output ports, as well as continuous spatiotemporal modes in each channel.
A natural framework for such an extension is provided by input-output theory~\cite{Gardiner1985, Kiilerich2019}.
In particular, actual quadrature measurements are implemented through balanced homodyne detection~\cite{Lvovsky2009, Tyc2004, Raymer1995}, and a quantitative assessment of the mode matching with the local oscillator is necessary to evaluate the performance of the ultrafast heralding discussed here.
Nevertheless, if a single dominant spatiotemporal mode is effectively selected both in the light--matter interaction and in the measurement of the outgoing light---for example, when a dominant Schmidt mode exists and mode matching with the local oscillator is high---then the single effective-mode description used here should remain a good approximation.

\section*{Acknowledgments}
This work was supported by JST FOREST (Grant No. JPMJFR2131) and JSPS KAKENHI (Grant Nos. JP25K17343 and JP26K21749).

\appendix
\renewcommand{\thesection}{Appendix~\Alph{section}}
\section{Backaction-induced unconditional macroscopic quantum-state generation at late times} \label{sec:extensions:unconditional}
We investigate the long-term dynamics of the coupled light--matter system and demonstrate that matter can be driven into a quantum state even without conditioning on optical measurement outcomes, due to the previously neglected backaction of matter on light.
Because this action--backaction process is rate-limited by the electron--photon coupling constant $g$, it proceeds on a slower timescale than the driving set by the effective field strength $g \sqrt{\langle \hat{n}\rangle}$ of strong-field quantum light.
Accordingly, relaxation and dephasing generally compete with this dynamics; however, in the present analysis, we restrict ourselves to the coherent limit.

\begin{figure}[t]
\centering
\includegraphics[width=\columnwidth]{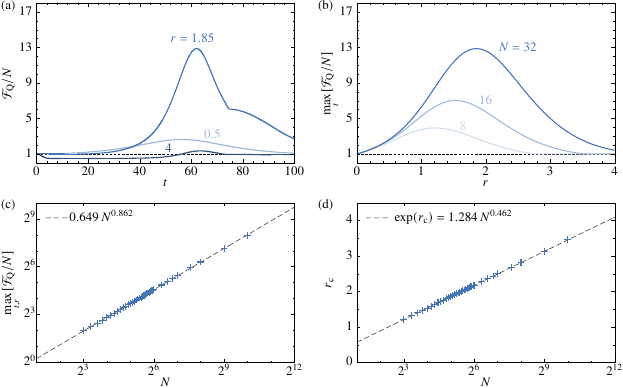}
\caption{
Late-time unconditional matter dynamics in the TC model~\eqref{eq:def:tc_hamiltonian} without the XFA.
The electron--photon coupling is fixed at $g=0.005$.
(a)~Time profile of the QFI density $\mathcal{F}_{\mathrm{Q}}/N$ calculated from $\hat{\rho}_{\mathrm{m}}$ [Eq.~\eqref{eq:def:unconditional}] for $N=32$.
The curves correspond to $r = 0.5$, $1.85$, and $4$.
(b)~Maximum QFI over time as a function of the squeezing parameter~$r$.
(c)~Maximum QFI over time and $r$ as a function of the particle number $N$.
(d)~Particle-number dependence of the optimal squeezing parameter $r_{\mathrm{c}}$ that maximizes the QFI.
}
\label{fig:ftrace_out}
\end{figure}
Figure~\ref{fig:ftrace_out} shows the dynamics of the unconditional matter density matrix $\hat{\rho}_{\mathrm{m}}$ [Eq.~\eqref{eq:def:unconditional}] calculated from the total-system wave function time-evolved under the TC model~\eqref{eq:def:tc_hamiltonian}.
The computational method and conditions are the same as those of Fig.~\ref{fig:qfull}, with the electron--photon coupling fixed at $g = 0.005$.
Figure~\ref{fig:ftrace_out}(a) shows the time profile of the QFI density $\mathcal{F}_{\mathrm{Q}}/N$ [Eq.~\eqref{eq:def:QFI}].
In the multiphoton regime $r = 4$ (dark blue curve), the QFI density does not exceed unity for $t \lesssim 40$, consistent with the unconditional XFA results shown as the red curve in Fig.~\ref{fig:qcoarse-grain}.
At later times, after once reaching the stationary value $(N-1)/[2(N+1)]$, the QFI density rises again, which can be interpreted as a manifestation of matter-to-light backaction.

With weaker squeezing and hence fewer photons, backaction influences the dynamics already in the early stage.
In the few-photon regime $r=0.5$ (light blue curve), the QFI density exhibits a peaked increase.
This dynamics can be understood as quantum-state transfer, in which the matter system absorbs all the energy initially contained in the light field, together with its squeezing~\cite{Kuzmich1997, Hald1999}; see Ref.~\cite{Imai2026} for the cat-state case.

For $N=32$, the QFI peak originating from this complete energy absorption reaches its maximum around $r=1.85$ (blue curve); the peak time scales as $t_{\mathrm{peak}} \propto 1/(g\sqrt{N})$ because of the collective coupling.
Figure~\ref{fig:ftrace_out}(b) shows the dependence of the time-maximized QFI density, $\max_t[ \mathcal{F}_{\mathrm{Q}}/N]$, on the squeezing parameter $r$.
Again, the QFI density is nonmonotonic in $r$ and exhibits a peak.
In particular, the decrease of $\max_t[ \mathcal{F}_{\mathrm{Q}}/N]$ for $r \gg 1$ is consistent with the XFA conclusion that the quantumness characterized by the QFI does not increase in the unconditional case.

Figure~\ref{fig:ftrace_out}(c) shows the particle-number dependence of this QFI-density peak.
The numerical data can be fitted by
\begin{equation}
\max_{r}\left[ \max_{t}[\mathcal{F}_{\mathrm{Q}}/N] \right] \approx 0.65\, N^{0.86}. \label{eq:full:QFI_vs_N}
\end{equation}
Although this does not reach the Heisenberg scaling $\mathcal{F}_{\mathrm{Q}}\propto N^2$, it suggests metrological sensitivity beyond the SQL ($\mathcal{F}_{\mathrm{Q}}\propto N$).
However, because relaxation and dephasing are neglected here, environmental effects should be carefully considered for this quantum-state formation in the long-time regime ($t \sim g^{-1}$).

Figure~\ref{fig:ftrace_out}(d) further shows the particle-number dependence of the optimal squeezing parameter $r_{\mathrm{c}}$ that maximizes the QFI density.
The numerical results are well fitted by
\begin{equation}
r_{\mathrm{c}} \approx 0.46\, \ln N + 0.26. \label{eq:full:squeezing_vs_N}
\end{equation}

\printbibliography
\end{document}